\documentclass[aip,preprint]{revtex4-1}
\usepackage{graphicx}
\usepackage{calc}
\usepackage{amsmath}
\usepackage{amssymb}

\draft 

\begin{document}

\title{Ab initio simulation of the universal properties of unitary Fermi gas in a harmonic trap}

\author{Yunuo Xiong}
\email{xiongyunuo@hbpu.edu.cn}
\affiliation{Center for Fundamental Physics and School of Mathematics and Physics, Hubei Polytechnic University, Huangshi 435003, China}

\author{Hongwei Xiong}
\email{xionghongwei@hbpu.edu.cn}

\affiliation{Center for Fundamental Physics and School of Mathematics and Physics, Hubei Polytechnic University, Huangshi 435003, China}

\date{\today}

\begin{abstract}
Chang and Bertsch [Phys. Rev. A 76, 021603(R) (2007)] proposed a simple formula for the ground state energy of a unitary Fermi gas in a harmonic trap, based on their Green's function Monte Carlo simulations of up to 22 fermions, combined with general assumptions about the universal thermodynamic behavior of the unitary Fermi gas. In this work, we perform the \textit{ab initio} simulations of the ground state energy of up to one hundred fermions using the fictitious identical particle method to overcome the Fermion sign problem, and we find that the formula proposed by Chang and Bertsch remains highly accurate. Since the number of fermions we simulate is much larger than that simulated by Chang and Bertsch when they proposed the formula, our work provides strong evidence for the universal validity of the formula. Our work demonstrates that fictitious identical particles provide a valuable tool for the \textit{ab initio} simulations of ultracold Fermi gases.
\end{abstract}

\maketitle

\section{Introduction}

For the unitary Fermi gas, the s-wave scattering length $a_s$ 
between fermions of different spins is divergent. Therefore, we expect this strongly correlated quantum system to exhibit simple universal behavior \cite{BertschO,Ho,Strinati,Thomas,Sagi,Ku,Horikoshi,LiX}. For the unitary Fermi gas in a three-dimensional periodic box, the chemical potential $\mu$ and the Fermi energy $E_F$ of the non-interacting Fermi gas have a simple relationship: $\mu=\xi_b E_F$, 
where $\xi_b$ is the Bertsch parameter. For the uniform unitary Fermi gas, extensive theoretical studies \cite{BlochRMP,Pollet,Carlson1,Astrakharchik,Haussmann,Burov2,BulgacU,LiuXJ,Forbes} have been carried out to investigate this universal behavior, and it has been confirmed by experiments \cite{Thomas,Sagi,Ku,Horikoshi} on ultracold Fermi gases.

For a unitary Fermi gas in a harmonic trap, universal behavior is still expected, and it is conjectured that the harmonic trap should be reflected in the universal behavior. For a unitary Fermi gas trapped in a three-dimensional harmonic trap with angular frequency $\omega$, Chang and Bertsch\cite{GFMC} proposed the following energy formula:
\begin{equation}
E=\xi_b^{1/2}\omega\left(\frac{(3N)^{4/3}}{4}+\frac{(3N)^{2/3}}{8}\right).
\label{Bertsch}
\end{equation}
The coefficient $\xi_b$ here is a universal constant similar to the Bertsch parameter. $N=N_\uparrow+N_\downarrow$ is the total number of fermions. $N_\uparrow$ and $N_\downarrow$  are the number of fermions with different spin states. This formula is proposed for the spin-balanced case, i.e. when $N_\uparrow=N_\downarrow$. Chang and Bertsch proposed the above formula after calculating from $N=2$ to $N=22$ fermions, and determined the coefficient $\xi_b=0.5$ by fitting.

Chang and Bertsch \cite{GFMC} used the Green's Function Monte Carlo method (GFMC) to simulate the ground state energy of fermions. Unfortunately, no GFMC simulation results have been seen for larger scale unitary Fermi gases with far more than 22 fermions. Since the above formula (\ref{Bertsch}) is tested for 22 or fewer fermions and the parameter $\xi_b$ is obtained by fitting, a rigorous test of this formula requires calculating the ground state energy of more fermions. If universal behavior does exist, then Eq. (\ref{Bertsch}) and the parameter $\xi_b$ should also hold for the case of far more than 22 fermions.

Due to the difficulty brought by the Fermion sign problem \cite{ceperley,troyer,WDM,Dornheim-PRE,Alex}, the \textit{ab initio} simulation of large-scale Fermi gases in harmonic traps is quite difficult. With the increase of the number of fermions, the computational cost of GFMC shows an exponential increase, which also faces the challenge of large-scale simulation. Recently, fictitious identical particle thermodynamics \cite{XiongFSP,Xiong-xi,Dornheim1,Dornheim2,Dornheim3,Dornheim4} has provided a new method to overcome the Fermion sign problem. In particular, the $\xi$-extrapolation method of fictitious identical particles has achieved unprecedented level of consistency with independent experimental observations in the \textit{ab initio} simulating warm dense matter in the National Ignition Facility, leading to a major breakthrough in path integral Monte Carlo (PIMC) \cite{Dornheim3}. Another breakthrough is the successful application of the $\xi$-extrapolation method by Dornheim et al. \cite{Dornheim4} to simulate warm dense hydrogen and beryllium, which provides a solution for the accurate \textit{ab initio} simulation of warm dense hydrogen at metallic density. In this work, we will utilize and develop the method of fictitious identical particles to simulate the ground state energy of Fermi systems, and calculate the ground state energy of unitary Fermi gases with up to 100 fermions. Our simulation results show that Eq. (\ref{Bertsch}) is highly valid for different fermion numbers N=20, 30, 40, 60, 70, 80, 100, thus verifying the universal behavior described by Eq. (\ref{Bertsch}). This work also confirms that fictitious identical particles have value in overcoming the Fermion sign problem and simulating the thermodynamic properties of ultracold fermionic gases.

The structure of this paper is as follows. In Sec. \ref{properties}, we briefly introduce fictitious identical particle thermodynamics and give two basic properties. In Sec. \ref{ideaS}, we introduce the zero-temperature interpolation method for inferring the ground state energy of Fermi systems based on fictitious identical particle thermodynamics. In Sec. \ref{application}, we apply the zero-temperature interpolation method to unitary Fermi gases, and illustrate the robustness of the zero-temperature interpolation method by numerical experiments. In Sec. \ref{unitary}, we calculate the ground state energy of unitary Fermi gases with different fermion numbers, and verify Eq. (\ref{Bertsch}) proposed by Chang and Bertsch about universal behavior. In Sec. \ref{summar}, we give a brief summary and discussion.

\section{Fictitious Identical Particle Thermodynamics and Several Basic Properties}
\label{properties}

To introduce the method for inferring the ground state energy of Fermi systems in this work, we first introduce several important properties that can be obtained from the partition function based on fictitious identical particles. In fictitious identical particle thermodynamics \cite{XiongFSP,Xiong-xi}, we introduce a real parameter $\xi$ to describe the fictitious identical particles. $\xi=1$ represents bosons, $\xi=-1$ represents fermions. $\xi$ is a real number that can vary continuously.

Fictitious identical particles with $\xi\neq \pm 1$ do not exist in the real world of elementary particles, but the thermodynamic properties of fictitious identical particles can be studied mathematically. We consider the following partition function:
\begin{equation}
Z(\beta)=Tr(e^{-\beta\hat H}).
\end{equation}
Here $\beta=1/k_BT$ with $k_B$ the Boltzmann constant and $T$ the temperature. The parametrized partition function for single-component fictitious identical particles with a parameter $\xi$ can be written as
\begin{equation}
Z(\xi,\beta)\sim\sum_{p\in S_N}\xi^{N_p}\int d\textbf{r}_1d\textbf{r}_2\cdots d\textbf{r}_N<p\{\textbf{r}\}|e^{-\beta\hat H}|\{\textbf{r}\}>.
\label{Xipartition}
\end{equation}
$S_N$ represents the set of $N!$ permutation operations denoted by $p$. The factor $\xi^{N_p}$ is due to the exchange effect of fictitious identical particles. $\xi=+1$ for bosonic partition function, while $\xi=-1$  for fermionic partition function. We adopt the symbol convention $0^0=1$ here so that distinguishable particles ($\xi=0$) can also be included in the above expression. In addition, $\{\textbf{r}\}$ denotes $\{\textbf{r}_1,\cdots,\textbf{r}_N\}$. $N_p$ is a number defined to be the minimum number of times for which pairs of indices must be interchanged to recover the original order $\{\textbf{r}\}$ from $p\{\textbf{r}\}$. 

\subsection{The first property}

Based on the above partition function (\ref{Xipartition}), the energy expression of fictitious identical particles is:
\begin{equation}
E(\xi,T)=-\frac{\partial \ln Z(\xi,\beta)}{\partial \beta}.
\end{equation} 
We can also further differentiate with respect to temperature to get
\begin{equation}
\frac{\partial E(\xi,T)}{\partial T}=\frac{1}{k_BT^2}(\left<E^2\right>-\left<E\right>^2).
\end{equation}
When $T\rightarrow 0$, the energy fluctuation also tends to 0. Due to the existence of the exponential factor $e^{-\beta\hat H}$, $(<E^2>-<E>^2)$ will approach 0 faster than $T^2$.
Therefore, we have
\begin{equation}
\left.\frac{\partial E(\xi,T)}{\partial T}\right|_{T=0}=0.
\label{ET0}
\end{equation}

The above expression can also be proved by another method. We set $S(\xi,T)$ as the entropy of fictitious identical particles. Therefore, we have
\begin{equation}
dE(\xi,T)=TdS(\xi,T).
\end{equation}
Therefore,
\begin{equation}
\frac{\partial E(\xi,T)}{\partial T}=T\frac{\partial S(\xi,T)}{\partial T}.
\end{equation}
When $T\rightarrow 0$, $\frac{\partial S(\xi,T)}{\partial T}$ must be a finite value, which proves Eq. (\ref{ET0}).

Using the partial derivative formula
\begin{equation}
\frac{\partial \xi(E,T)}{\partial T}=-\frac{\partial E(\xi,T)/\partial T}{\partial E(\xi,T)/\partial \xi},
\label{partial}
\end{equation}
we also have
\begin{equation}
\left.\frac{\partial \xi(E,T)}{\partial T}\right|_{T=0}=0.
\label{exactE}
\end{equation}
It is this exact relationship that led us to propose the constant energy semi-extrapolation method \cite{Xiong-xi},  and it will play a role in this paper.

Eq. (\ref{ET0}) can be further extended. Let's imagine another physical quantity $p(\xi,T,\textbf C)$. Here $\textbf C$ represents an independent variable, for example, if we consider the density distribution, then $\textbf C$ represents the spatial coordinate $\textbf r$. Since the energy $E$ is a function of $\xi$ and $T$, we can also express this physical quantity as $p(\xi,T,\textbf C)$. Therefore, we have
\begin{equation}
\frac{\partial p(\xi,E(\xi,T),\textbf C)}{\partial T}=\frac{\partial p(\xi,E(\xi,T),\textbf C)}{\partial E}\frac{\partial E(\xi,T)}{\partial T}.
\end{equation}
Therefore, we have
\begin{equation}
\left.\frac{\partial p(\xi,T,\textbf C)}{\partial T}\right|_{T=0}=0.
\end{equation}
Further using the partial derivative formula similar to Eq. (\ref{partial}), we can get:
\begin{equation}
\left.\frac{\partial \xi(p,T,\textbf C)}{\partial T}\right|_{T=0}=0.
\end{equation}

\subsection{The second property}

We now consider the ground-state properties of the bosonic sector $0\leq \xi\leq 1$ at $T=0$. At $T=0$, the quantum state of the Bose system ($\xi=1$) is in the ground state and can be described by a single many-body wave function $\Psi_g({\textbf r}_1,\cdots,{\textbf r}_i,\cdots,{\textbf r}_j,\cdots,{\textbf r}_N; \xi=1)$. Here, the coordinates ${\textbf r}_i$ and ${\textbf r}_j$  must satisfy exchange symmetry. Since exchange symmetry can be effectively seen as an attractive interaction, we have the property that the ground state energy $E_g(\xi)$ is monotonically decreasing or remains constant in the bosonic sector. Obviously, for $\xi=0$, the many-body wave function $\Psi_g({\textbf r}_1,\cdots,{\textbf r}_i,\cdots,{\textbf r}_j,\cdots,{\textbf r}_N; \xi=1)$ can also be used to describe the system. Therefore, we find that $E_g$ of the Bose system is actually the ground state energy of distinguishable particles. Considering that $\xi$ is a continuous real number and the monotonic behavior of the energy with respect to $\xi$, we obtain the following property of the ground state energy:
\begin{equation}
E_g(0\leq \xi\leq 1,T=0)= const.
\end{equation}

Similarly, for other physical quantities, we also have the property that $p(0\leq\xi\leq 1,T=0,\textbf C)$ is $\xi$-independent. Considering the antisymmetry of the wave function of the Fermi system under the exchange of identical particles, we cannot extend the above properties to the fermionic sector.

\subsection{The third property}

For the sake of simplicity, we use the symbol $\int$ to replace$\int d\textbf{r}_1d\textbf{r}_2\cdots d\textbf{r}_N$. Therefore, the expression of $E(\xi,T)$ is:
\begin{equation}
E(\xi,T)=\frac{\sum_{p\in S_N}\xi^{N_p}\int <p\{\textbf{r}\}|\hat He^{-\beta\hat H}|\{\textbf{r}\}>}{\sum_{p\in S_N}\xi^{N_p}\int <p\{\textbf{r}\}|e^{-\beta\hat H}|\{\textbf{r}\}>}.
\end{equation}

We can also rewrite the summation expressions in the numerator and denominator above as:
\begin{equation}
E(\xi,T)=\frac{\sum_{N_p=0}^{N-1}\xi^{N_p}\sum_p'\int <p\{\textbf{r}\}|\hat He^{-\beta\hat H}|\{\textbf{r}\}>}{\sum_{N_p=0}^{N-1}\xi^{N_p}\sum_p'\int <p\{\textbf{r}\}|e^{-\beta\hat H}|\{\textbf{r}\}>}.
\end{equation}
Here, $\sum_p'$ represents the summation over all permutations for a given $N_p$. Therefore, we have
\[
\frac{\partial E(\xi,T)}{\partial \xi}=\frac{(\sum_{N_p=1}^{N-1}N_p\xi^{N_p-1}\sum_p'\int <p\{\textbf{r}\}|\hat He^{-\beta\hat H}|\{\textbf{r}\}>)(\sum_{N_p=0}^{N-1}\xi^{N_p}\sum_p'\int <p\{\textbf{r}\}|e^{-\beta\hat H}|\{\textbf{r}\}>)}{(\sum_{N_p=0}^{N-1}\xi^{N_p}\sum_p'\int <p\{\textbf{r}\}|e^{-\beta\hat H}|\{\textbf{r}\}>)^2}
\]

\begin{equation}
-\frac{(\sum_{N_p=0}^{N-1}\xi^{N_p}\sum_p'\int <p\{\textbf{r}\}|\hat He^{-\beta\hat H}|\{\textbf{r}\}>)(\sum_{N_p=1}^{N-1}N_p\xi^{N_p-1}\sum_p'\int <p\{\textbf{r}\}|e^{-\beta\hat H}|\{\textbf{r}\}>)}{(\sum_{N_p=0}^{N-1}\xi^{N_p}\sum_p'\int <p\{\textbf{r}\}|e^{-\beta\hat H}|\{\textbf{r}\}>)^2}.
\end{equation}

When $\xi\rightarrow 0$, we have
\[
\left.\frac{\partial E(\xi,T)}{\partial \xi}\right|_{\xi=0}=\frac{(\sum_p^{N_p=1}\int <p\{\textbf{r}\}|\hat He^{-\beta\hat H}|\{\textbf{r}\}>)(\sum_p^{N_p=0}\int <p\{\textbf{r}\}|e^{-\beta\hat H}|\{\textbf{r}\}>)}{(\sum_p^{N_p=0}\int <p\{\textbf{r}\}|e^{-\beta\hat H}|\{\textbf{r}\}>)^2}
\]

\begin{equation}
-\frac{(\sum_p^{N_p=0}\int <p\{\textbf{r}\}|\hat He^{-\beta\hat H}|\{\textbf{r}\}>)(\sum_p^{N_p=1}\int <p\{\textbf{r}\}|e^{-\beta\hat H}|\{\textbf{r}\}>)}{(\sum_p^{N_p=0}\int <p\{\textbf{r}\}|e^{-\beta\hat H}|\{\textbf{r}\}>)^2}.
\end{equation}
When $\beta\rightarrow\infty$ or $T\rightarrow 0$, for the case of $N>>1$, we have
\begin{equation}
\sum_p^{N_p=0}\int <p\{\textbf{r}\}|\hat He^{-\beta\hat H}|\{\textbf{r}\}>\backsimeq 
\sum_p^{N_p=1}\int <p\{\textbf{r}\}|\hat He^{-\beta\hat H}|\{\textbf{r}\}>,
\end{equation}

\begin{equation}
\sum_p^{N_p=0}\int <p\{\textbf{r}\}|e^{-\beta\hat H}|\{\textbf{r}\}>\backsimeq 
\sum_p^{N_p=1}\int <p\{\textbf{r}\}|e^{-\beta\hat H}|\{\textbf{r}\}>.
\end{equation}

The reason why the above two relations hold is that there is only one permutation for $N_p=1$ compared to $N_p=0$; when $N>>1$, these two cases will be more similar when performing PIMD (path integral molecular dynamics) or PIMC (path integral Monte Carlo) sampling. In addition, when $T\rightarrow 0$, both samplings are performed for the ground state, and the difference between the two cases will be further reduced. Of course, for the case of $\xi\neq 0$ and $T$ significantly non-zero, the above analysis no longer holds.

Therefore, we have
\begin{equation}
\left.\frac{\partial E(\xi,T)}{\partial \xi}\right|_{\xi=0,T=0}\backsimeq 0.
\label{exact}
\end{equation}

\begin{figure}[htbp]
\begin{center}
\includegraphics[scale=0.3]{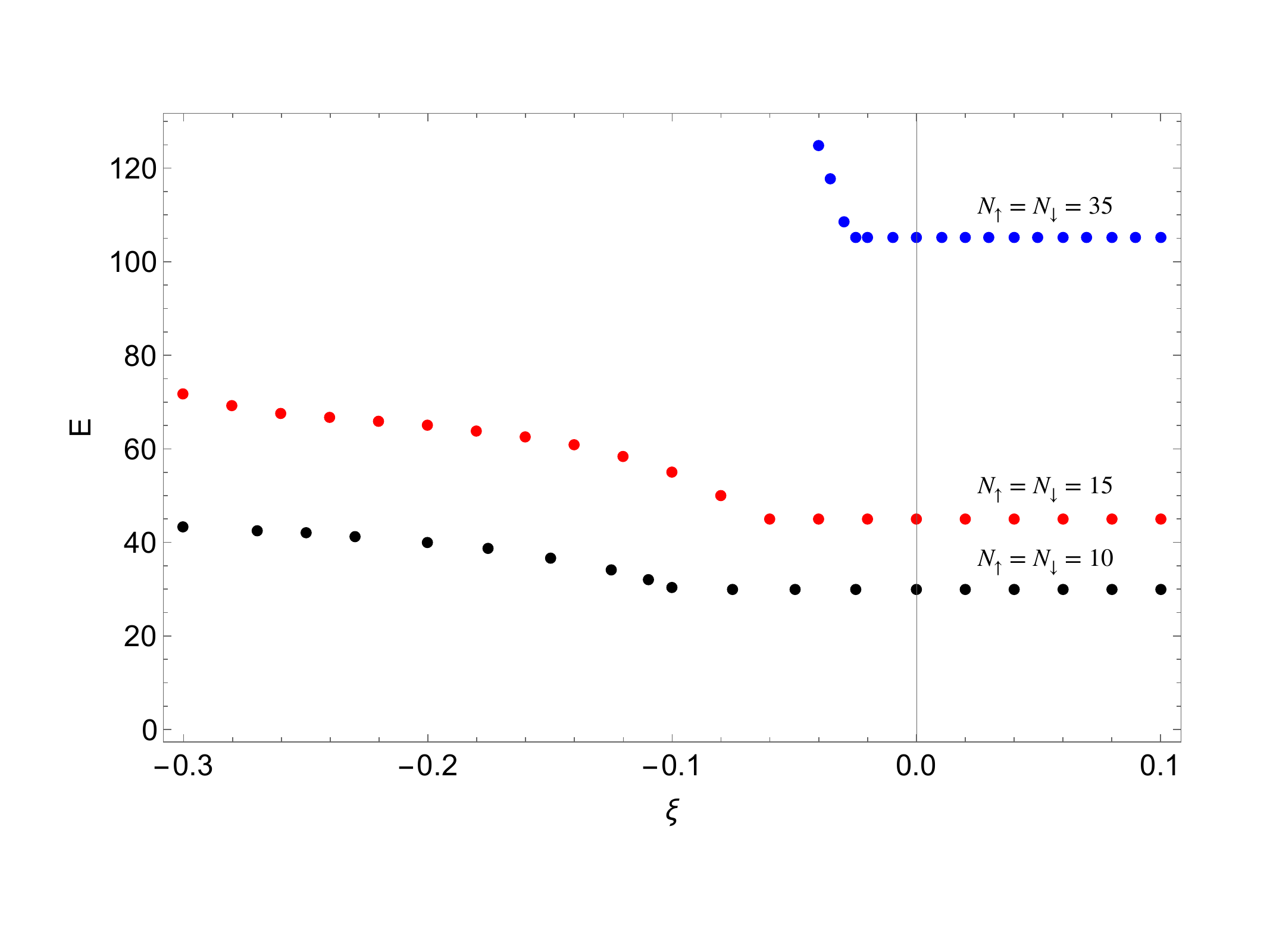}
\caption{In the convention $\hbar=\omega=1$, we give the relationship between the ground state energy of two-component fictitious identical particles in a three-dimensional isotropic harmonic  trap and $\xi$.}
\label{idealenergy}
\end{center}
\end{figure}

For non-interacting fictitious identical particles in a three-dimensional harmonic trap, we can obtain exact results, which provide a test of the above properties. In Fig. \ref{idealenergy}, we show the ground state energy at $T=0$ for two-component fictitious identical particles with different spin states in the cases of $N_\uparrow=N_\downarrow=10,15,35$. We note that this example satisfies properties 2 and 3.

For a physical quantity $p(\xi,T,\textbf C)$, similar to the above derivation, we also have
\begin{equation}
\left.\frac{\partial p(\xi,T,\textbf C)}{\partial \xi}\right|_{\xi=0,T=0}\backsimeq 0.
\label{exactP}
\end{equation}

The above three properties are obtained for fictitious identical particles of a single component, and also apply to the thermodynamics of fictitious identical particles of two components.

\section{General Idea of Inferring the Ground State Energy of Fermi Systems Using Zero-Temperature Interpolation Method}
\label{ideaS}

\begin{figure}[htbp]
\begin{center}
\includegraphics[scale=0.3]{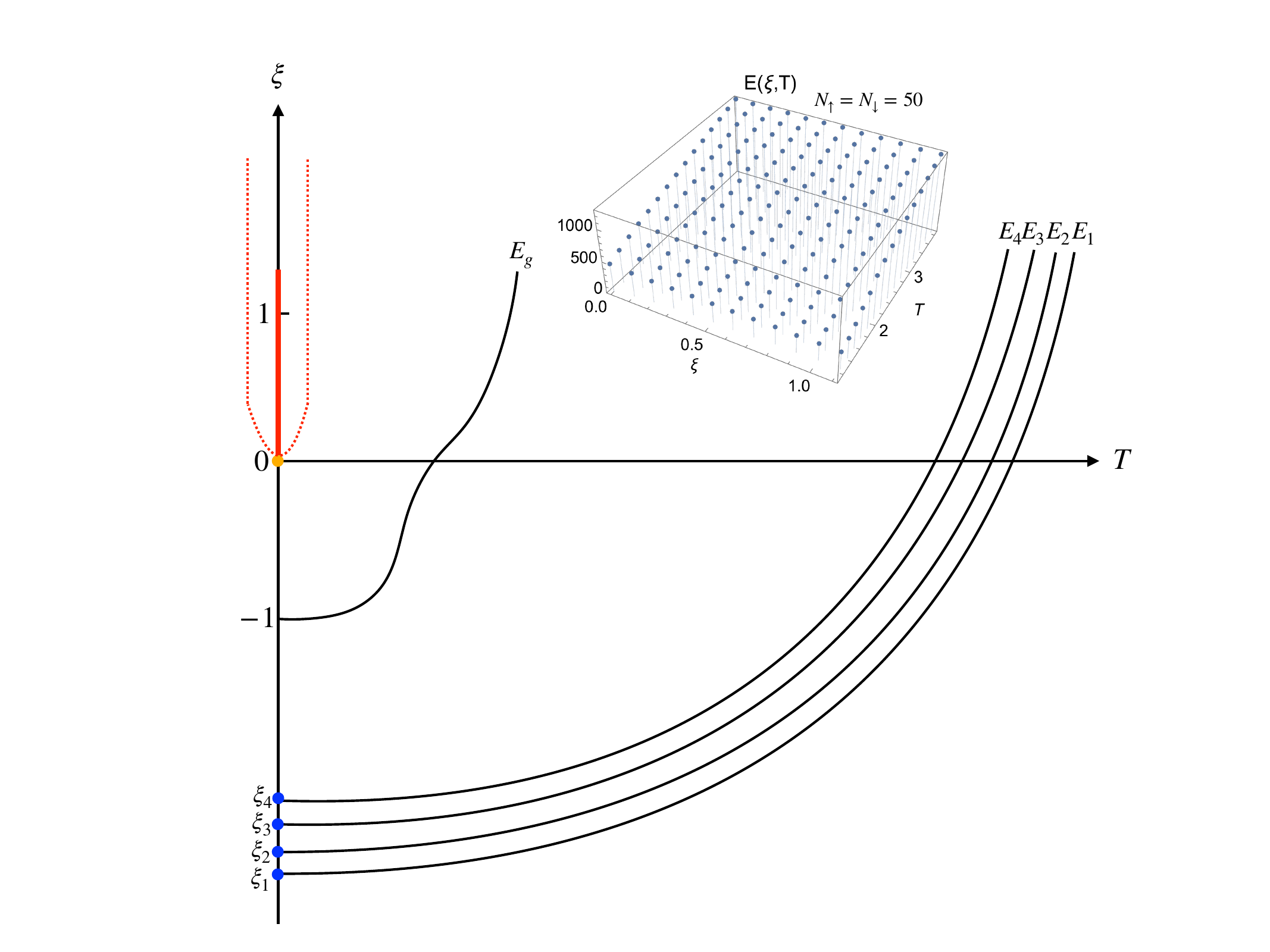}
\caption{Fictitious identical particle thermodynamics extends the thermodynamic properties that depend on temperature to a two-dimensional abstract space of $(\xi,T)$, providing a variety of rich means for inferring the thermodynamic properties of Fermi systems. The inset of the figure shows the energy simulation results of 100 particles in the bosonic sector in the unitary limit. $E_g,E_1,E_2,E_3,E_4$ represent the energy contours in the $\xi-T$ coordinate system. $E_g$  represents the ground state energy of the fermions. For the case of attractive interactions, the energy contour of $E_g$  is usually complicated and it is difficult to directly extrapolate the curves of the fermionic sector based on the data in the bosonic sector. For the case of higher energies, the energy contours are much simpler, and it is thus hopeful to reliably extrapolate to the fermionic sector after using the exact condition (\ref{exactE}). The red solid line represents that the energy of the bosonic sector is the same at $T=0$ according to Property 2 in Sec. \ref{properties}, which terminates at $\xi=0$ and no longer holds for $\xi<0$.}
\label{curve}
\end{center}
\end{figure}

The general idea of inferring the thermodynamic properties of Fermi systems using fictitious identical particle thermodynamics can be briefly illustrated by Fig. \ref{curve}. In the inset of Fig. \ref{curve}, we show the energy simulation results for different $\xi$ and temperatures based on PIMD simulations \cite{HirshPIMD,HirshImprove,Deuterium,Xiong-spinor,Xiong-Momentum,Xiong-magnetic,Xiong-anyon,Xiong-Green,Xiong-Hubbard} for a unitary gas in the bosonic sector ($\xi\geq 0$). In the bosonic sector, there is no Fermion sign problem, and we present here the energy simulation results for 100 particles in the unitary limit. Our task is to infer the thermodynamic properties of the Fermi system from the large number of simulation results in the bosonic sector. For the case of non-interacting or repulsive interactions, previous studies have shown that isothermal extrapolation \cite{XiongFSP,Dornheim1,Dornheim2} can accurately simulate the energy of the Fermi system in the weak or moderate quantum degeneracy regime, while constant energy semi-extrapolation \cite{Xiong-xi} can hopefully give the Fermi energy at high quantum degeneracy.

In Fig. \ref{curve}, we show several energy contours $\xi_E(T)$. We are particularly interested in the energy contour corresponding to the ground state energy $E_g$ of the Fermi system. Our previous studies have shown that for the case of repulsive interactions, the energy contours are generally simple, so that the constant energy semi-extrapolation method \cite{Xiong-xi,Xiong-Hubbard} can be used to infer the ground state energy of the Fermi system. Compared with isothermal extrapolation, one advantage of constant energy semi-extrapolation is that there exists an exact condition $\left.\frac{\partial \xi(E,T)}{\partial T}\right|_{T=0}$ that can be adopted. However, for the case of attractive interactions, especially when there is a finite temperature phase transition of the Fermi system, the energy contour corresponding to the ground state energy $E_g$ may be complicated, such as the typical energy contour we show in Fig. \ref{curve} for the case of attractive interactions. Later, we will prove the existence of this shape. In the region $\xi\geq 0$, even if we know a series of data $\{\xi_j,T_j\}$ of the energy contour of $E_g$, it is not easy to find a sufficiently accurate energy contour expression to confirm that the energy contour of $E_g$  intersects exactly with the point $(\xi=-1,T=0)$ and is perpendicular to the $\xi$ axis.

How to solve this tricky problem?  In Fig. \ref{curve}, the red solid line represents that the energy in the bosonic sector is the same at $T=0$, which prohibits us from deducing the ground state energy of the Fermi system via the isothermal expansion method based on the energy of the bosonic sector. In Fig. \ref{curve}, the red solid line is enclosed by dashed lines. In the theory of functions, we often encounter such situations where it is difficult to directly extend the real line with $\xi>0$ to 
$\xi=-1$ along $T=0$ by analytic continuation. A commonly used approach is to bypass the region enclosed by the red dashed lines and obtain the ground state energy of the Fermi system from the thermodynamic properties of the bosonic sector by means of a clever analytic continuation method. This work will demonstrate how to successfully achieve this goal based on this insightful idea.

For the energy contour, in the high temperature region, the exchange effect of identical particles will weaken, so the energy contour corresponding to high energy will be simpler, especially there will be no strange behavior near $\xi=0$. Since the length of the energy contour at this time is also larger than that of the low temperature energy contour, the curvature will also be smaller, which makes it easier to accurately characterize. This is the essential reason why isothermal extrapolation is applicable to medium and high temperatures \cite{XiongFSP,Dornheim1,Dornheim2,Dornheim3,Dornheim4}. In Fig. \ref{curve}, the energy contours corresponding to $E_1,E_2,E_3,E_4$  represent this simpler situation. Based on this consideration, we have hope to accurately know the ground state energy of the fictitious identical particles corresponding to a $\xi_j$ less than $-1$ at $T=0$, which is $E_j$. We use the blue dots in Fig. \ref{curve} to represent.

\begin{figure}[htbp]
\begin{center}
\includegraphics[scale=0.3]{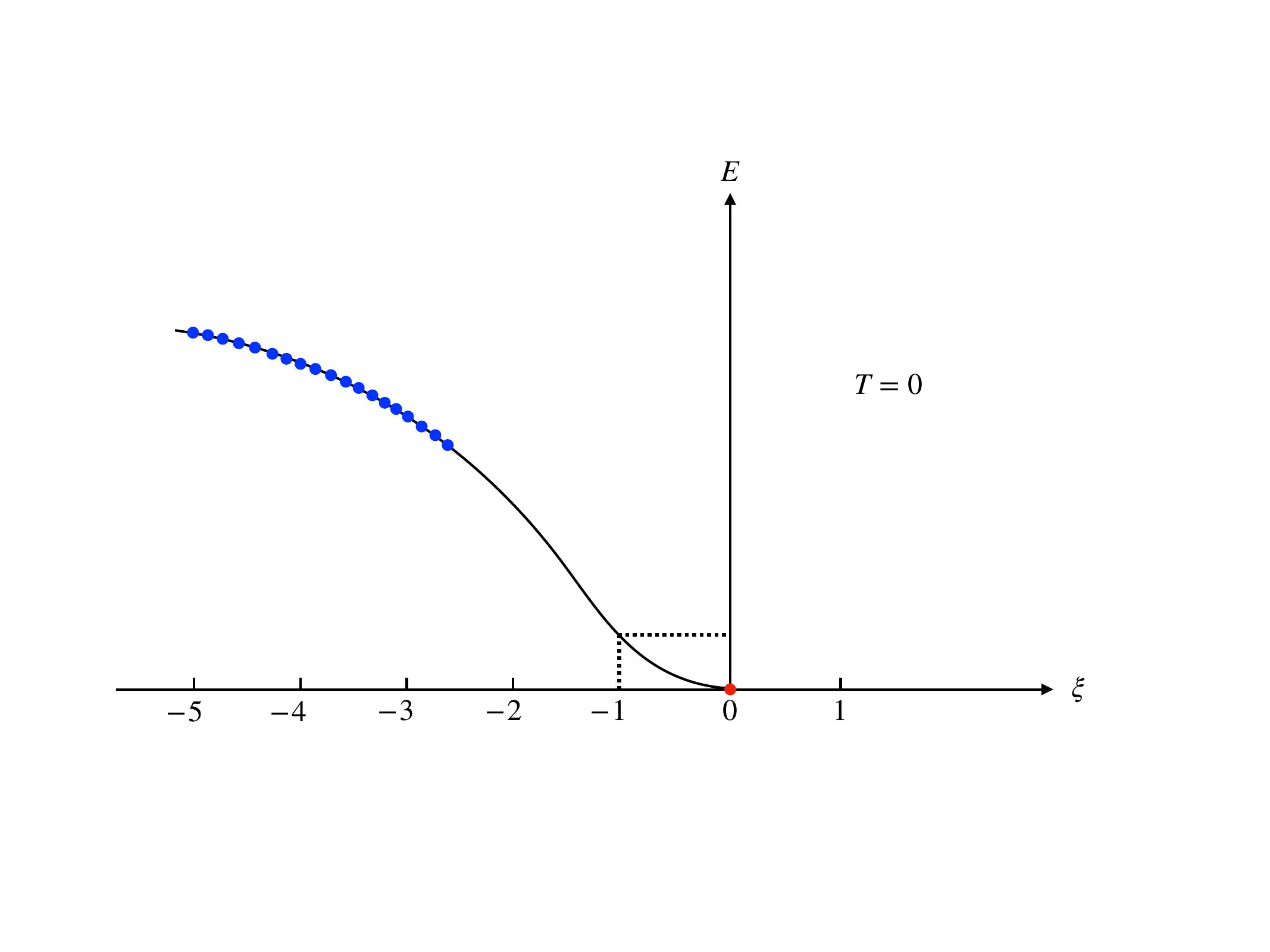}
\caption{The figure shows the idea of inferring the ground state energy of the Fermi system using the interpolation method at zero temperature. The blue dots represent the values of $\xi$ at zero temperature inferred from the constant energy semi-extrapolation at different energies. The red dot represents the ground state energy at $\xi=0$. Further combining the condition (\ref{exact}), we have hope to accurately fit the black curve $E(\xi)$, and thus infer the ground state energy of the Fermi system $E(\xi=-1)$ based on this interpolation method.}
\label{idea}
\end{center}
\end{figure}

In addition, we can in principle accurately simulate the ground state energy $E_d$ of distinguishable particles at $(\xi=0,T=0)$. We use the red dot in Fig. \ref{idea} to represent. In this case, we have the following data at $T=0$:
\[
\{\xi=0,E_d\},\{\xi_1<-1,E_1\},\cdots,\{\xi_j<-1,E_j\}
\]
In Fig. \ref{idea}, we use red dot and blue dots to represent these data respectively. We now want to fit these data to obtain the function $E_g(\xi)$, and thus obtain $E_g(\xi=-1)$, which is the ground state energy of the Fermi system. When choosing the fitting function, we also have the condition (\ref{exact}) to adopt, which is greatly helpful for finding a reasonable fitting function. We note that in this fitting process, we know both the data for $\xi<-1$ and $\xi>-1$. With the further help of condition (\ref{exact}), we have hope to accurately obtain the ground state energy of the Fermi system through the interpolation method. As a comparison, in the previous finite temperature isothermal extrapolation method, an extrapolation method was used. The reliability of this isothermal extrapolation method depends on the fact that at medium and high temperatures, $E(\xi)$ can be accurately described by a quadratic function under isothermal conditions \cite{XiongFSP,Dornheim1,Dornheim2,Dornheim3,Dornheim4}.

In a sense, what we propose here is a zero temperature interpolation method. In previous work, isothermal extrapolation was only suitable for medium and high temperatures \cite{XiongFSP,Dornheim1,Dornheim2,Dornheim3,Dornheim4}. In these works, isothermal extrapolation first calculates the energy data for $\xi\geq 0$ at the same temperature, and then obtains the fitting function $E(\xi)$, and extrapolates it to the energy of the Fermi system $E(\xi=-1)$. This extrapolation ability will fail when $T$ is close to $0$, because when $T$ is very small, $E(\xi\geq 0)$ only weakly depends on $\xi$, while there should be a strong dependence when $\xi<0$. In the current work, for the isothermal interpolation method at $T=0$, we perform the fitting in the region $\xi\leq 0$. The reason why we have hope to develop a variety of methods to infer the thermodynamic properties of Fermi systems is that in fictitious identical particle thermodynamics, we extend the usual thermodynamic properties with respect to temperature to a two-dimensional abstract space of $\xi$ and $T$.

\section{Application of the Zero-Temperature Interpolation Method to the unitary Fermi Gas}
\label{application}

It is worth pointing out that the \textit{ab initio} simulations can be viewed as a numerical experiment. We need to verify the validity of the method by comparing it with existing reliable benchmarks or actual experimental phenomena. Before various numerical experiments are obtained, we often cannot arbitrarily draw conclusions on whether the method is valid based on the strictest mathematical analysis; and it is also difficult to determine the accuracy of the new method without numerical experiments, but through pure mathematical analysis. Due to the difficulty of the Fermion sign problem, it is difficult for us to directly use PIMD/PIMC to simulate in the fermionic sector $\xi<0$ to obtain the accurate behavior of the general $(\xi,T)$ shown in Fig. \ref{curve} in the fermionic sector. At this time, only the comparison of numerical experimental results and physical reasonableness analysis can give the verification criteria for whether the method is valid.

For the zero-temperature interpolation method proposed above, the unitary Fermi gas in the three-dimensional harmonic trap \cite{Thomas,Sagi,Ku,Horikoshi,LiX,GFMC,Gilbreth,latticeMC,Mukherjee,Carlson,FNDMC} provides an excellent opportunity for verification and application. Chang and Bertsch's numerical simulation \cite{GFMC} of the unitary Fermi gas provides a benchmark for testing the zero-temperature interpolation method, while applying the zero-temperature interpolation method to the case of more fermions can in turn test the validity of Eq. (\ref{Bertsch}).

In a three-dimensional harmonic trap with attractive interaction, the dimensionless Hamiltonian operator is
\begin{equation}
\hat{H}=-\frac{1}{2}\sum_{j=1}^N\Delta_j+\frac{1}{2}\sum_{j=1}^N\textbf{x}_j^2-\sum_{j=1}^{N_\uparrow}\sum_{j'=1}^{N_\downarrow} \frac{ V}{w^2} e^{-|{\textbf x}_j-{\textbf x}_{j'}|^2/w^2}.
\end{equation}
We use the usual convention $\hbar=k_B=m=\omega=1$ here.
The Gaussian interaction in the above Hamiltonian operator can easily adjust the s-wave scattering length $a_s$. To assure the situation of divergent scattering length, the parameter in this paper is chosen as \cite{GaussianInt}
\begin{equation}
w=0.01,~~~~~V=2.684.
\end{equation}
For this choice of $w$, the attractive interaction is short-range and much smaller than the average distance between the fermions studied here, which satisfies the condition of the unitary limit $w<<l<<a_s$. Here $l$ represents the average distance between particles.

\begin{figure}[htbp]
\begin{center}
\includegraphics[scale=0.3]{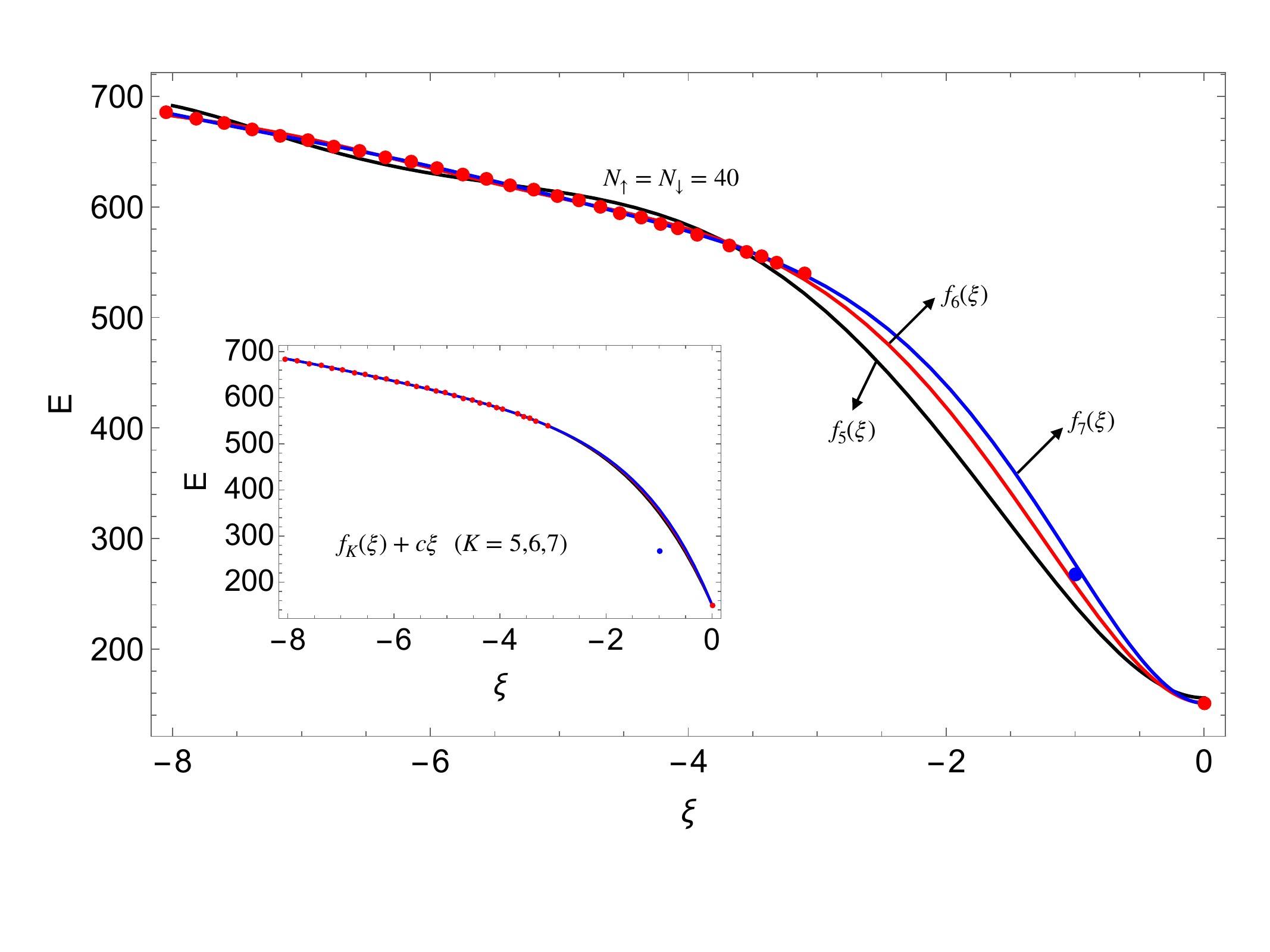}
\caption{The figure shows the fitting results of the zero-temperature interpolation method for the unitary Fermi gas with $N_\uparrow=N_\downarrow=40$. The red dots are the energy data given by the constant energy semi-extrapolation and the ground state energy of $\xi=0$. These data represented by red dots are used to determine the undetermined coefficients $a$ and $b_j$ in the fitting function $f_K(\xi)$. The blue dot is the energy predicted by Eq. (\ref{Bertsch}). In the figure, we use three fitting functions $f_K(\xi)$ (K=5,6,7). The inset of the figure shows the results using the fitting function $f_K(\xi)+c\xi$ (K=5,6,7), which has an obvious deviation from the blue dot.}
\label{4040}
\end{center}
\end{figure}

We take $N_\uparrow=N_\downarrow=40$ as an example to analyze specifically below. In Fig. \ref{4040}, we use red dots to represent the data $E_j(\xi_j,T=0)$ obtained by the energy contour of the fitting function $\xi=a+bT^2$
  for $\xi<-1$, as well as the ground state energy of $(\xi=0,T=0)$ simulated by PIMD. According to these data, we can use a suitable fitting function to fit. Considering condition (\ref{exact}), we choose the following fitting function:
\begin{equation}
f_K(\xi)=a+\sum_{j=2}^Kb_j\xi^j.
\end{equation}
Since this fitting function does not contain a linear term, it satisfies the condition $\frac{df_K(\xi)}{d\xi}|_{\xi=0}=0$.

In Fig. \ref{4040}, we show the results of three fitting functions $f_K(\xi)$ (K=5,6,7). We note that there is a clear underfitting of $f_5(\xi)$. When $K<5$, there is a more obvious underfitting, which is not shown in the figure. $f_6(\xi)$ and $f_7(\xi)$ both achieve satisfactory fitting and interpolation, which indicates the robustness of the fitting and shows that there is no need to worry about the overfitting problem in the current example. The reason why we don't have to worry about overfitting is that the fitting here is actually a kind of interpolation method. If we extrapolate to the regions of $\xi<-8$ and $\xi>0$, there will be obvious deviations. Fortunately, in the current method, we are concerned with the ground state energy of fermions at $\xi=-1$. This is precisely the advantage that interpolation can bring. The blue dot in Fig. \ref{4040} represent the ground state energy given by Eq. (\ref{Bertsch}). The average ground state energy of fermions given by the two successful fitting functions is 267.49 (of which the result given by $f_6(\xi)$ is 258.04, and the result given by $f_7(\xi)$ is 276.94), and the ground state energy given by Eq. (\ref{Bertsch}) is 267.07. The energy fluctuation caused by the choice of two different fitting functions $f_6(\xi)$ and $f_7(\xi)$ is 3\%. In the inset of Fig. \ref{4040}, we also give the fitting results without considering the condition  
$\frac{d f_K(\xi)}{d \xi}|_{\xi=0}=0$. At this time, the fitting function we adopt is $f_K(\xi)+a\xi$ (K=5,6,7). Although these three fitting functions are highly coincident due to the characteristics of the interpolation method, the ground state energy of fermions at $\xi=-1$ and the prediction of Eq. (\ref{Bertsch}) have obvious deviations.

As we all know, the interpolation method has significant advantages over the extrapolation method, especially when we do not know the exact expression of the function in advance and simply use polynomial functions to fit. Above, we took the example of $N_\uparrow=N_\downarrow=40$ to reveal the reliability of the zero-temperature interpolation method. However, we still face a question that needs to be verified. In the fitting of the zero-temperature interpolation method, the energy of $(\xi=0,T=0)$ can be achieved with high accuracy, but for the energy data of $(\xi<-1,T=0)$, we can still reasonably doubt whether the constant energy semi-extrapolation is accurate enough. The following numerical experiments show that we can still reliably infer the ground state energy of the Fermi system without the need for constant energy semi-extrapolation to obtain very accurate results.

\begin{figure}[htbp]
\begin{center}
\includegraphics[scale=0.3]{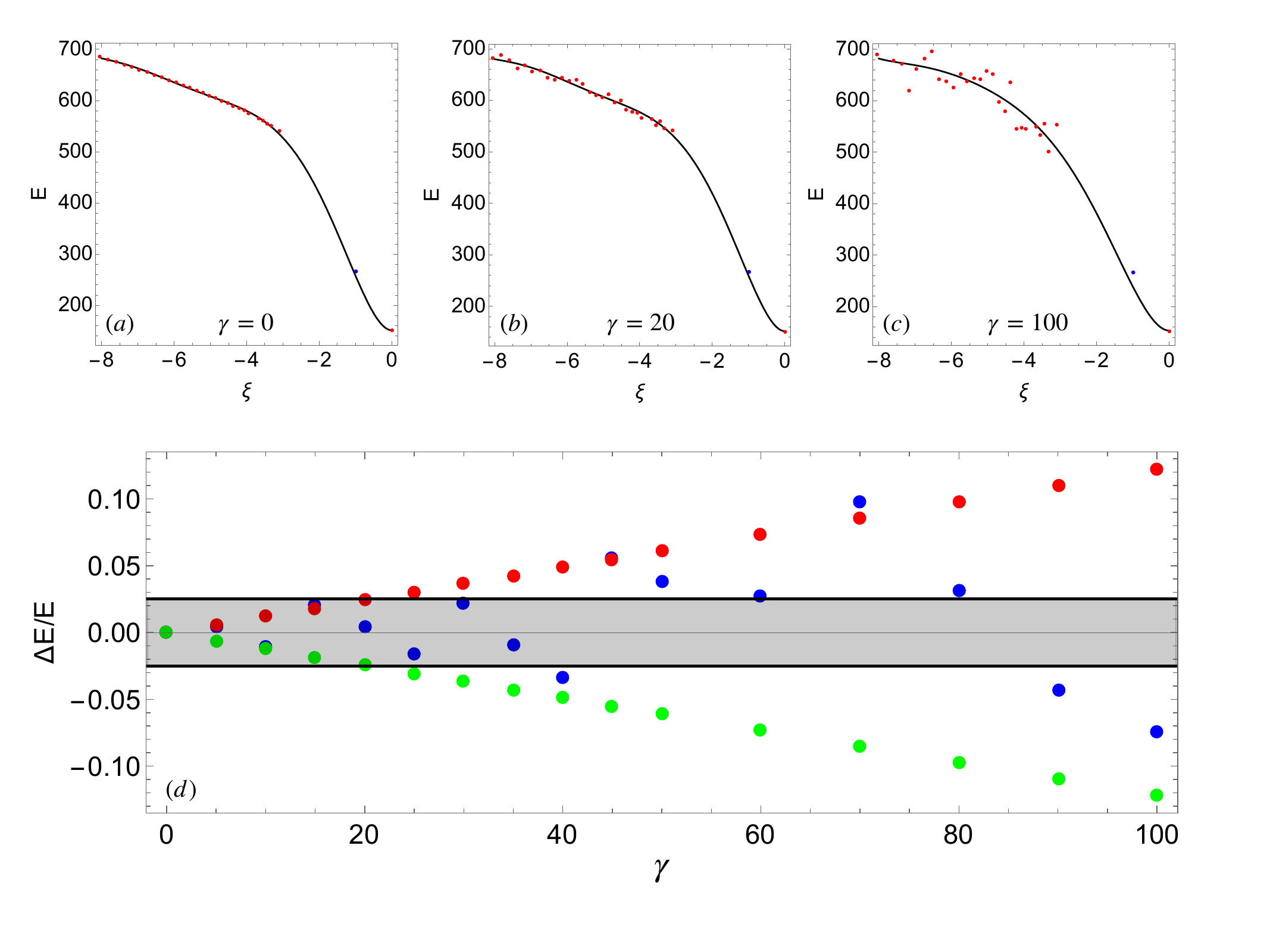}
\caption{Figs. (a)-(c) show the energy data and fitting curves after adding different noise strengths $\gamma$ for the case of $N_\uparrow=N_\downarrow=40$. The blue dots in Fig. (d) represent the relative deviation $\Delta E/E$ of the energy obtained by the zero-temperature interpolation method after adding uniform noise between $[-\frac{\gamma}{2},\frac{\gamma}{2}]$ to the energy data $E_j$ in $\xi<-1$. The red dots (green dots) in Fig. (d) represent the relative deviation $\Delta E/E$ of the energy obtained by the zero-temperature interpolation method after adding the same $\gamma$ (-$\gamma$) to the energy data $E_j$ in $\xi<-1$.}
\label{noise}
\end{center}
\end{figure}

For the energy data $\{\xi_j,E_j\}$ of $\xi<-1$ and $T=0$ with $N_\uparrow=N_\downarrow=40$, we artificially add random noise $n_j$ to $E_j$. Here $n_j$ is uniform noise in the range $[-\frac{\gamma}{2},\frac{\gamma}{2}]$, and $\gamma$ represents the noise strength. In Figs. \ref{noise}(a)-(c), we use red dots to give the energy data for $\gamma=0,20,100$. We note that at $\gamma=20$, the noise is already quite significant. After adding noise, the black solid line in the figure is the fitting result of $f_6(\xi)$, and the blue dot represents the prediction of Eq. (\ref{Bertsch}). In Fig. \ref{noise}(d), we use blue dots to represent the relative deviation $\Delta E/E$ after applying different noise strengths $\gamma$, compared to the energy without noise. We find that when $\gamma<40$, the fluctuation amplitude of $\Delta E/E$ is only 2.5\%. This confirms the noise resistance of the zero-temperature interpolation method in predicting the ground state energy of fermions, and the reason is of course because it is an interpolation fitting method. 

We also consider the case where there is a systematic deviation in the constant energy semi-extrapolation. For this case, we artificially add the same $\gamma$ to $E_j$ in $\xi<-1$ before fitting. The red dots in the figure give the relationship between $\Delta E/E$ and $\gamma$, and the green dots give the relationship between $\Delta E/E$ and $-\gamma$. We also note that the zero-temperature interpolation method has good robustness to the possible systematic deviation in the constant energy semi-extrapolation. One of the reasons for this robustness is that we know the accurate energy of $(\xi=0,T=0)$ and there is condition (\ref{exact}) in the fitting.

\section{Universal Behavior of the Ground State Energy of the Unitary Fermi Gas}
\label{unitary}

\begin{figure}[htbp]
\begin{center}
\includegraphics[scale=0.5]{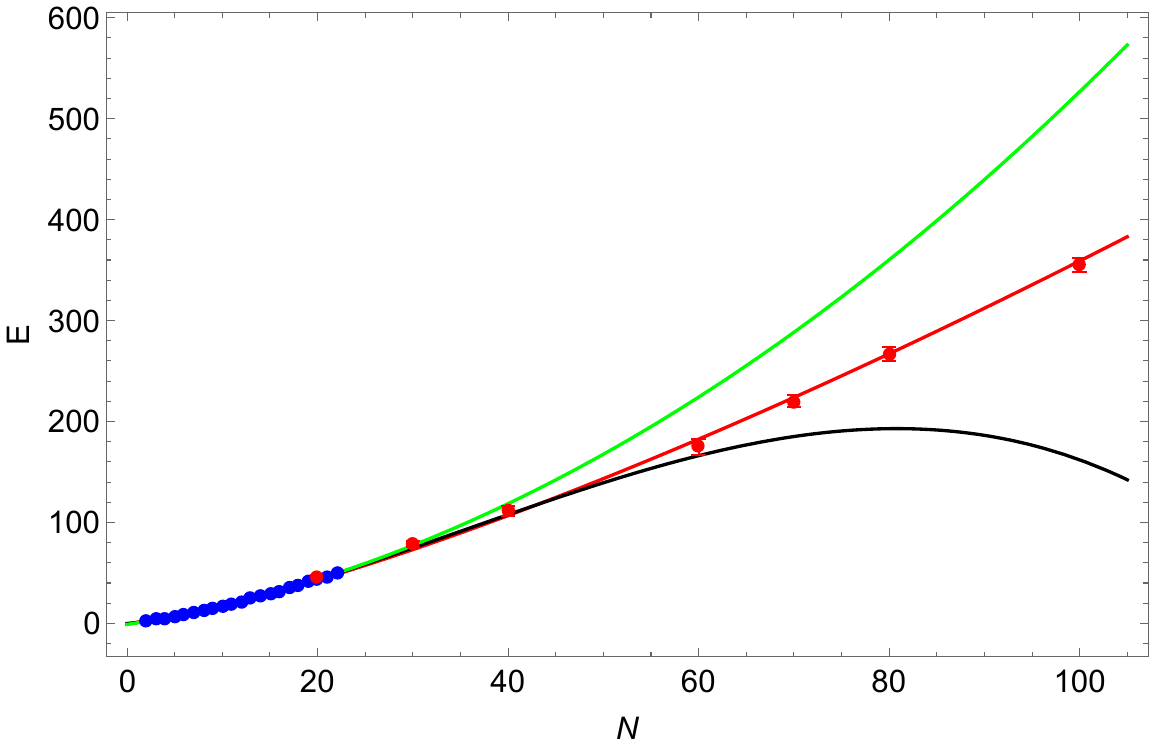}
\caption{The blue dots in the figure are the ground state energies of different fermion numbers obtained by Chang and Bertsch using GFMC simulation. The red line is the energy given by Eq. (\ref{Bertsch}) obtained by fitting the blue dots. The red dots are the ground state energies inferred by the zero-temperature interpolation method based on fictitious identical particles in this work, which are highly consistent with the extrapolation results of the red line. In order to fully reflect the extrapolation value of Eq. (\ref{Bertsch}), the green line and the black line in the figure are the extrapolation curves obtained by fitting the blue dot data with $a+b N+c N^2$  and $a+b N+c N^2+d N^3$, respectively. Since these two fitting functions do not take into account the universal behavior of the unitary Fermi gas, they quickly fail completely as the number of fermions increases. This comparison fully demonstrates the general validity of Eq. (\ref{Bertsch}).}
\label{energy}
\end{center}
\end{figure}

In order to further systematically test the universal thermodynamic behavior of the unitary Fermi gas, we here study the different cases of N=20, 30, 40, 60, 70, 80, 100. In Fig. \ref{energy}, the red dots represent the ground state energy of fermions given by the zero-temperature interpolation method. In Fig. \ref{energy}, the blue dots are the results obtained by Chang and Bertsch based on GFMC simulation \cite{GFMC}. The red line in the figure is the ground state energy of fermions given by Eq. (\ref{Bertsch}). We note that in the whole region of different fermion numbers, the red dots given by the zero-temperature interpolation method coincide with the red line within the range of fluctuation.

In GFMC, it is necessary to construct a multi-particle wave function of fermion pairs in advance, while in our method, it is not necessary to make any prior assumptions about the properties of the quantum system. 
In the article by Chang and Bertsch \cite{GFMC}, when fitting the parameter $\xi_b$ based on Eq. (\ref{Bertsch}), only the GFMC simulation data with $N\leq 22$ were used. However, due to the plausibility of the conjectured universal behavior, this fitting formula is valid for all cases we simulated, especially when $N=100$. This provides strong evidence for the validity of Eq. (\ref{Bertsch}). In Fig. \ref{energy}, we also use green and black lines to give the fitting results of the polynomial function based on the data with $N\leq 22$ without assuming the universal behavior. We note that when $N>40$, there is an obvious difference between the predictions given by the zero-temperature interpolation method and Eq. (\ref{Bertsch}). The fundamental reason why the extrapolation given by the red line is highly consistent with the result of the zero-temperature interpolation method is the simple universal behavior of the unitary Fermi gas, and Chang and Bertsch correctly discovered the analytical expression of this universal behavior.

\begin{figure}[htbp]
\begin{center}
\includegraphics[scale=0.3]{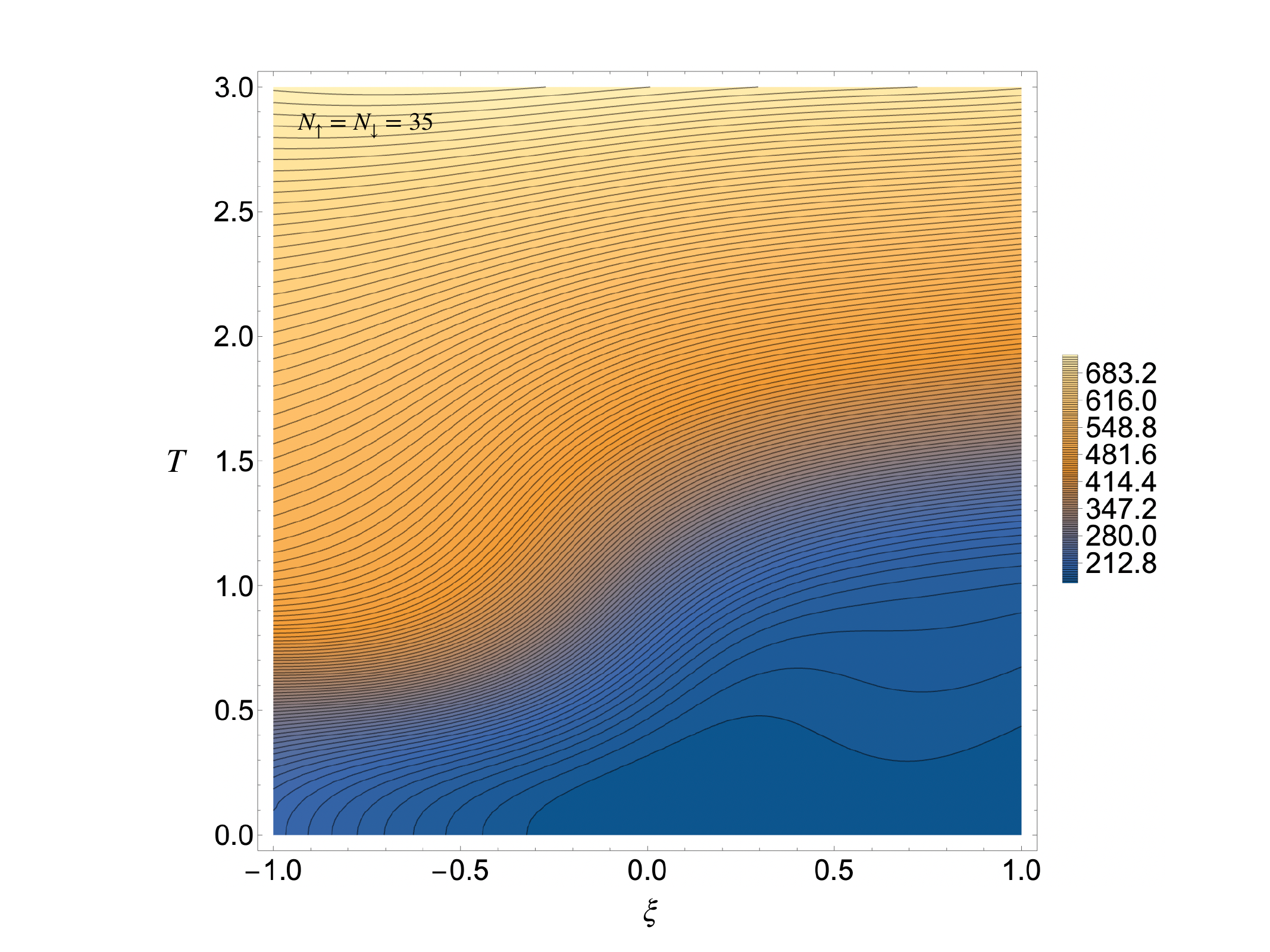}
\caption{The figure shows the contour map of the energy function $E(\xi,T)$ of $N_\uparrow=N_\downarrow=35$ obtained by physics-informed neural networks.}
\label{3535energy}
\end{center}
\end{figure}

One may wonder whether the shape of the constant energy curve of the ground state energy of the Fermi system caused by the attractive interaction in Fig. \ref{curve} is true. In the paper \cite{XiongPINNs}, we proposed a method to simulate the finite temperature phase transition of the Fermi system using physics-informed neural networks (PINNs) \cite{PINN1,PINN2,PINN3}. Based on PINNs, we can reveal the general relationship between energy and $(\xi,T)$. After completing the training by also using the ground state energy of the Fermi system obtained in this work as the input data of PINNs, we show the contour map of the energy function $E(\xi,T)$ of $N_\uparrow=N_\downarrow=35$ obtained by PINNs in Fig. \ref{3535energy}. We note that for the ground state of the Fermi system, the shape of the constant energy curve is similar to that shown in Fig. \ref{curve}. This situation will lead to the difficulty of constant energy semi-extrapolation based on data with $\xi>0$. We also note that when $E>550$, the constant energy curve becomes simpler, so we can use the constant energy semi-extrapolation method to obtain the specific value of $\xi<-1$ corresponding to this energy, and thus bypass the trouble caused by the complexity of the constant energy curve when $E<550$. In the article \cite{XiongPINNs}, we also give the PINNs simulation of the specific heat and phase transition of the Fermi system, and the critical temperature obtained from the specific heat is consistent with the experimental observation of the unitary Fermi gas in the harmonic trap.

\section{Conclusion and discussion}
\label{summar}

In summary, we have calculated the ground state energy of the unitary Fermi gas based on the fictitious identical particle thermodynamics. Through the \textit{ab initio} simulations with up to 100 fermions, we confirmed the universal thermodynamic behavior of the unitary Fermi gas in the harmonic trap proposed by Chang and Bertsch \cite{GFMC}. At the same time, this confirmation also proves the applicability of fictitious identical particle thermodynamics to the unitary Fermi gas.

It should be emphasized that although fictitious identical particle thermodynamics has the advantage of computational efficiency due to the essential overcoming of the Fermion sign problem when simulating large-scale Fermi systems, the method for inferring the thermodynamic properties of Fermi systems is still under development \cite{Dornheim1,Dornheim2}. At this time, the benchmarks provided by methods such as GFMC are of great value to the development of fictitious identical particle thermodynamics. In fact, in establishing the isothermal extrapolation method, Dornheim's use \cite{Dornheim-PRE} of direct PIMC \cite{CeperleyBook,Feynman,Tuckerman,Fosdick,Jordan,Herman,barker,Morita,CeperleyRMP,Burov1,Burov1b} to perform the \textit{ab initio} simulations of various quantum systems played a key benchmark role. The unitary Fermi gas can also play an important role in this regard, because the universal thermodynamic behavior proposed by Chang and Bertsch can also provide a benchmark for large-scale Fermi systems.

So far, for the \textit{ab initio} simulation of energy, we have established the isothermal extrapolation method \cite{XiongFSP}, the constant energy semi-extrapolation method \cite{Xiong-xi}, PINNs \cite{XiongPINNs}, and the zero-temperature interpolation method here based on fictitious identical particle thermodynamics. This work shows that these tools have important prospects for the \textit{ab initio} simulation of ultracold Fermi gases. Future research can be carried out on spin-imbalanced \cite{Schunck,Shin} ultracold Fermi gases, which are more challenging for other methods that overcome the Fermion sign problem. In fictitious identical particle thermodynamics, the number of fermions in different spin states does not lead to additional difficulties. It is a pity that there is a great lack of the \textit{ab initio} benchmarks for the ground state energy of spin-imbalanced ultracold Fermi gases, so conducting the \textit{ab initio} simulations and revealing new quantum phenomena for related experiments is a future research direction.

\begin{acknowledgments}
This work is partly supported by the National Natural Science Foundation of China under grant numbers 11175246, and 11334001. This work has received funding from Hubei Polytechnic University.
\end{acknowledgments}




\begin{thebibliography}{10}


\bibitem{BertschO} T. Papenbrock and G. F. Bertsch, Pairing in low-density Fermi gases, Phys. Rev. C \textbf{59}, 2052 (1999).

\bibitem{Ho} T. L. Ho, Universal thermodynamics of degenerate quantum gases in the unitarity limit, Phys. Rev. Lett. \textbf{92}, 090402 (2004).


\bibitem{Strinati} G. C. Strinati, P. Pieri, G. Röpke, P. Schuck, and M. Urban, The BCS–BEC crossover: From ultra-cold Fermi gases to nuclear systems, Phys. Rep. \textbf{738}, 1-76 (2018).


\bibitem{Thomas} J. E. Thomas, J. Kinast,  and A. Turlapov, Virial theorem and universality in a unitary Fermi gas, Phys. Rev. Lett. \textbf{95}, 120402 (2005).

\bibitem{Sagi} Y. Sagi, T. E. Drake, R. Paudel, and D. S. Jin, Measurement of the homogeneous contact of a unitary Fermi gas, Phys. Rev. Lett. \textbf{109}, 220402 (2012).

\bibitem{Ku} M. J. Ku, A. T. Sommer, L. W. Cheuk, and M. W. Zwierlein, Revealing the superfluid lambda transition in the universal thermodynamics of a unitary Fermi gas, Science \textbf{335}, 563-567 (2012).

\bibitem{Horikoshi} M. Horikoshi, S. Nakajima, M. Ueda, and T. Mukaiyama,  Measurement of universal thermodynamic functions for a unitary Fermi gas, Science \textbf{327}, 442-445 (2010).

\bibitem{LiX} X. Li, S. Wang, X. Luo, Y. Y. Zhou, K. Xie, H. C. Shen, ...,  J. W. Pan, Observation and quantification of the pseudogap in unitary Fermi gases, Nature \textbf{626}, 288-293 (2024).

\bibitem{BlochRMP} I. Bloch, J. Dalibard, W. Zwerger, Many-body physics with ultracold gases, Rev. Mod. Phys. \textbf{80}, 885 (2008).

\bibitem{Pollet} L. Pollet, Recent developments in quantum Monte Carlo simulations with applications for cold gases, Rep. Mod. Phys. \textbf{75}, 094501 (2012).

\bibitem{Carlson1} J. Carlson, S. Y.  Chang, V. R.  Pandharipande, K. E.  Schmidt,  Superfluid Fermi gases with large scattering length, Phys. Rev. Lett. \textbf{91}, 050401 (2003).

\bibitem{Astrakharchik} G. E. Astrakharchik, J. Boronat, J. Casulleras, S. Giorgini, Equation of state of a Fermi gas in the BEC-BCS crossover: A quantum Monte Carlo study, Phys. Rev. Lett.  \textbf{93}, 200404 (2004).

\bibitem{Haussmann} R. Haussmann, W. Rantner, W. Cerrito, W.  Zwerger,  Thermodynamics of the BCS-BEC crossover, Phys. Rev. A \textbf{75}, 023610 (2007).

 \bibitem{Burov2} E. Burovski, N. V. Prokof’ev, B. V. Svistunov, and M. Troyer, Critical Temperature Curve in BEC-BCS Crossover, \text{Phys. Rev. Lett.}~  \textbf{101}, 090402 (2008).
 
\bibitem{BulgacU} A. Bulgac, J. E. Drut, P.  Magierski, Quantum Monte Carlo simulations of the BCS-BEC crossover at finite temperature, Phys. Rev. A \textbf{78}, 023625 (2008).

\bibitem{LiuXJ} X. J. Liu, H. Hu, P. D. Drummond, Virial expansion for a strongly correlated Fermi gas, Phys. Rev. Lett. \textbf{102}, 160401 (2009).

\bibitem{Forbes} M. M. Forbes, S. Gandolfi, A.  Gezerlis, Resonantly interacting fermions in a box, Phys. Rev. Lett. \textbf{106}, 235303 (2011).

\bibitem{GFMC} S. Y. Chang and G. F. Bertsch, Unitary Fermi gas in a harmonic trap, Phys. Rev. A \textbf{76}, 021603 (2007).


\bibitem{ceperley} D. M. Ceperley, Path Integral Monte Carlo Methods for Fermions, in \textit{Monte Carlo and Molecular Dynamics of Condensed Matter Systems}, edited by K. Binder and G. Ciccotti (Editrice Compositori, Bologna, Italy, 1996).

\bibitem{troyer} M.~Troyer and U. J.~Wiese, 
Computational Complexity and Fundamental Limitations to Fermionic Quantum Monte Carlo Simulations,
{\text{Phys. Rev. Lett.} \textbf{94}, 170201} (2005).

\bibitem{WDM} T. Dornheim, S. Groth, and M. Bonitz, The uniform electron gas at warm dense matter conditions, Phys. Rep \textbf{744}, 1 (2018).

\bibitem{Dornheim-PRE} T.~Dornheim, 
The Fermion sign problem in path integral Monte Carlo simulations: quantum dots, ultracold atoms, and warm dense matter,
\text{Phys. Rev. E}~\textbf{100}, 023307 (2019).

\bibitem{Alex} A. Alexandru, G. Basar, P. F. Bedaque, and N. C. Warrington, 
Complex paths around the sign problem,
Rev. Mod. Phys. \textbf{94}, 015006 (2022).

\bibitem{XiongFSP} Y. Xiong and H. Xiong, On the thermodynamic properties of fictitious identical particles and the application to fermion sign problem, J. Chem. Phys. \textbf{157}, 094112 (2022).  

\bibitem{Xiong-xi} Y. Xiong and H. Xiong, On the thermodynamics of fermions at any temperature based on parametrized partition function, Phys. Rev. E  \textbf{107}, 055308 (2023).

\bibitem{Dornheim1} T. Dornheim, P. Tolias, S. Groth, Z. A. Moldabekov, J. Vorberger, and B. Hirshberg, Fermionic physics from \textit{ab initio} path integral Monte Carlo simulations of fictitious identical particles,  J. Chem. Phys. \textbf{159}, 164113 (2023).

\bibitem{Dornheim2} T. Dornheim, S. Schwalbe, Z. A.  Moldabekov, J. Vorberger, and P. Tolias, \textit{Ab initio} path integral Monte Carlo simulations of the uniform electron gas on large length scales, J. Phys. Chem. Lett. \textbf{15}, 1305 (2024).

\bibitem{Dornheim3} T. Dornheim, T. D\"oppner, P. Tolias,
M. P. B\"ohme, L. B. Fletcher, Th. Gawne, F. R. Graziani, D. Kraus, M. J. MacDonald, Zh. A. Moldabekov,
S. Schwalbe, D. O. Gericke, and J. Vorberger, Unraveling electronic correlations in warm dense quantum plasmas, arXiv:2402.19113 (2024).

\bibitem{Dornheim4} T. Dornheim, S. Schwalbe, M. P. B\"ohme, Z. A. Moldabekov, J. Vorberger, and P. Tolias, Ab initio path integral Monte Carlo simulations of warm dense two-component systems without fixed nodes: structural properties, arXiv: 2403.01979 (2024).




\bibitem{HirshPIMD} B. Hirshberg, V. Rizzi, and M. Parrinello, Path integral molecular dynamics for bosons, Proc. Natl. Acad. Sci. USA \textbf{116}, 21445 (2019).

\bibitem{HirshImprove} Y. M. Y. Feldman and B. Hirshberg, Quadratic Scaling Bosonic Path Integral Molecular Dynamics, J. Chem. Phys. \textbf{159}, 154107 (2023).

 \bibitem{Deuterium}   C. W. Myung, B. Hirshberg, and M. Parrinello, Prediction of a supersolid phase in high-pressure deuterium, \text{Phys. Rev. Lett.} \textbf{128}, 045301 (2022).

\bibitem{Xiong-spinor} Y. Yu, S. Liu, H. Xiong, and Y. Xiong, Path integral molecular dynamics for thermodynamics and Green's function of ultracold spinor bosons, J. Chem. Phys. \textbf{157}, 064110 (2022).

\bibitem{Xiong-Momentum} Y. Xiong and  H. Xiong, 
Numerical calculation of Green's function and momentum distribution for spin-polarized fermions by path integral molecular dynamics, 
J. Chem. Phys. \textbf{156}, 204117 (2022).

\bibitem{Xiong-magnetic} Y. Xiong and H. Xiong, Path integral and winding number in singular magnetic field, Eur. Phys. J. Plus \textbf{137}, 550 (2022).

\bibitem{Xiong-anyon} Y. Xiong and H. Xiong, Path integral molecular dynamics for anyons, bosons, and fermions, Phys. Rev. E \textbf{106}, 025309 (2022).

\bibitem{Xiong-Green} Y. Xiong and H. Xiong, Path integral molecular dynamics simulations for Green’s function in a system of identical bosons, J. Chem. Phys. \textbf{156}, 134112 (2022).

\bibitem{Xiong-Hubbard} Y. Xiong, S. Liu and H. Xiong, Quadratic scaling path integral molecular dynamics for fictitious identical particles and its application to fermion systems, arXiv: 2401.00274 (2024).

\bibitem{Gilbreth} C. N. Gilbreth and Y. Alhassid, Pair condensation in a finite trapped Fermi gas,  Phys. Rev. A  \textbf{88},  063643 (2013).

\bibitem{latticeMC} M. G. Endres, D. B. Kaplan, J. W. Lee,  and A. N. Nicholson, Lattice Monte Carlo calculations for unitary fermions in a harmonic trap, Phys. Rev. A \textbf{84}, 043644 (2011).

\bibitem{Mukherjee} A. Mukherjee and Y. Alhassid, Configuration-interaction Monte Carlo method and its application to the trapped unitary Fermi gas, Phys. Rev. A \textbf{88},  053622 (2013).  

\bibitem{Carlson} J. Carlson and S. Gandolfi, Predicting energies of small clusters from the inhomogeneous unitary Fermi gas, Phy. Rev. A \textbf{90},  011601 (2014). 

\bibitem{FNDMC} D. Blume, J. von Stecher, and C. H. Greene, Universal properties of a trapped two-component Fermi gas at unitarity, Phys. Rev. Lett. \textbf{99}, 233201 (2007).





\bibitem{GaussianInt} P. Jeszenszki, A. Y. Cherny, and J. Brand, The s-wave scattering length of a Gaussian potential, Phys. Rev. A \textbf{97}, 042708 (2018).

\bibitem{XiongPINNs} Y. Xiong and H. Xiong, 
Ab initio simulations of the thermodynamic properties and phase transition of Fermi systems based on fictitious identical particles and physics-informed neural networks, arXiv: 2402.07231 (2024).

\bibitem{PINN1} G. E. Karniadakis, I. G. Kevrekidis, L. Lu,  P. Perdikaris, S. Wang, L.  Yang, Physics-informed machine learning, Nat. Rev. Phys. \textbf{3}, 422-440 (2021).

\bibitem{PINN2} S. Cuomo, V. S. Di Cola, F. Giampaolo, G. Rozza, M. Raissi, F.  Piccialli, Scientific machine learning through physics–informed neural networks: Where we are and what’s next, J. Sci. Comput. \textbf{92}, 88 (2022).

\bibitem{PINN3} M. Raissi, P. Perdikaris, G. E. Karniadakis, Physics-informed neural networks: A deep learning framework for solving forward and inverse problems involving nonlinear partial differential equations, J. Comput. Phys. \textbf{378}, 686-707 (2019).



\bibitem{CeperleyBook} R. M. Martin, L. Reining, and D. M. Ceperley, \textit{Interacting Electrons: Theory and Computational Approaches} (Cambridge University Press, Cambridge, UK, 2016).

\bibitem{Feynman} R. P. Feynman and R. H.  Albert,  \textit{Quantum mechanics and path integrals}  (McGraw-Hill, New York, 1965).

\bibitem{Tuckerman} M. E.~Tuckerman, \textit{Statistical mechanics: theory and molecular simulation} (Oxford University, New York, 2010).

\bibitem{Fosdick} L. D. Fosdick and H. F. Jordan, Path-integral calculation of the two-particle Slater sum for He$^4$, Phys. Rev. \textbf{143}, 58–66 (1966).

\bibitem{Jordan} H. F. Jordan and L. D. Fosdick, Three-particle effects in the pair distribution function for He4 gas, Phys. Rev. \textbf{171}, 128–149 (1968).

\bibitem{Herman} M. F. Herman, E. J. Bruskin, and B. J. Berne, On path integral Monte Carlo simulations, J. Chem. Phys. \textbf{76}, 5150–5155 (1982).

\bibitem{barker} J. A.~Barker, A quantum-statistical Monte Carlo method; path integrals with boundary conditions, J. Chem. Phys. \textbf{70}, 2914 (1979).

\bibitem{Morita} T. Morita, Solution of the Bloch Equation for Many-Particle Systems in Terms of the Path Integral, J. Phys. Soc. Japan. \textbf{35}, 980 (1973).

\bibitem{CeperleyRMP} D. M. Ceperley, Path integrals in the theory of condensed helium, Rev. Mod. Phys. \textbf{67}, 279 (1995).

\bibitem{Burov1} M. Boninsegni, N. V. Prokof’ev, and B. V. Svistunov, Worm Algorithm for Continuous-Space Path Integral Monte Carlo Simulations, \text{Phys. Rev. Lett.} ~\textbf{96}, 070601 (2006). 

\bibitem{Burov1b} M. Boninsegni, N. V. Prokof’ev, and B. V. Svistunov, Worm algorithm and diagrammatic Monte Carlo: A new approach to continuous-space path integral Monte Carlo simulations, Phys. Rev. E \textbf{74}, 036701 (2006).


\bibitem{Schunck} C. H. Schunck, Y. Shin, A. Schirotzek, M. W. Zwierlein, W.  Ketterle,  Pairing without superfluidity: The ground state of an imbalanced Fermi mixture, Science \textbf{316}, 867-870 (2007).


\bibitem{Shin} Y. I. Shin, C. H. Schunck, A. Schirotzek, W. Ketterle, Phase diagram of a two-component Fermi gas with resonant interactions, Nature \textbf{451}, 689-693 (2008).


\end{thebibliography}
\end{document}